\documentclass{PoS}
\pdfoutput=1

\usepackage{amsmath,bbm,graphicx}

\newcommand\ev[1]{\left\langle{#1}\right\rangle}
\newcommand{\eps}{\varepsilon}
\newcommand{\ha}{\hat\alpha}
\newcommand{\hm}{\hat m}
\newcommand{\hmu}{\hat\mu}
\newcommand{\Nf}{{N_f}}
\newcommand{\ph}{\bigl\langle e^{2i\theta} \bigr\rangle}
\newcommand{\phs}{\bigl\langle e^{2i\theta}_s \bigr\rangle}
\newcommand{\I}[1]{{\cal H}^s_{\nu,#1}(\ha,\hm)} 

\DeclareMathOperator\tr{tr}

\title{Random matrix analysis of the QCD sign problem\thanks{Supported
    by the German Research Foundation (DFG).}}

\ShortTitle{Random matrix analysis of the QCD sign problem}

\author{Jacques Bloch and \speaker{Tilo Wettig}\\
  Institute for Theoretical Physics, University of Regensburg, 93040
  Regensburg, Germany\\ 
  E-mail: \email{jacques.bloch@physik.uni-regensburg.de},
  \email{tilo.wettig@physik.uni-regensburg.de}}

\abstract{The severity of the sign problem in lattice QCD at nonzero
  baryon density is measured by the average phase of the fermion
  determinant. Motivated by the equivalence of chiral random matrix
  theory and QCD to leading order in the epsilon regime, we compute
  the phase of the fermion determinant for general topology in random
  matrix theory as a function of the quark chemical potential and the
  quark mass. We find that the sign problem becomes milder with
  increasing topological charge.  The analytic predictions are
  verified by detailed numerical random matrix simulations.}

\FullConference{The XXVII International Symposium on Lattice Field Theory\\
		 July 26-31, 2009\\
		 Peking University, Beijing, China}

\begin{document}

\section{Introduction}

The fermion sign problem is a major obstacle for unquenched lattice
QCD simulations at nonzero density, see \cite{lat09:PdF} for a review
at this conference.  The severity of the sign problem is measured by
the average phase of the fermion determinant.  Let us denote the Dirac
operator by $D(m;\mu)$, where $m$ and $\mu$ are the quark mass and the
quark chemical potential, respectively, and write $\det\,
D(m;\mu)=re^{i\theta}\in\mathbb{C}$.  Then the average phase of the
two-flavor theory is given by $\ev{e^{2i\theta}}$.  It was shown in
\cite{Splittorff:2006fu} that the average phase is nonzero for
$\mu<m_\pi/2$, whereas for $\mu>m_\pi/2$ it is exponentially
suppressed in $\mu^2F^2V$.  Here, $m_\pi$ is the pion mass, $V$ is the
volume, and $F$ and $\Sigma$ (see \eqref{eq:scaling} below) are the
familiar low-energy constants in the effective chiral Lagrangian.

The microscopic regime of QCD (corresponding to the leading order in
the $\eps$-expansion) is defined by the requirement that the scaling
variables
\begin{align}
  \hm=mV\Sigma\quad\text{and}\quad\hmu^2=\mu^2F^2V 
  \label{eq:scaling}
\end{align}
are kept fixed in the $V\to\infty$ limit.  In this regime exact
analytical results for QCD can be derived using chiral random matrix
theory (RMT).  In \cite{Splittorff:2007ck} the average phase was
computed for zero topology from RMT.  An interesting question (e.g.,
for fixed-topology simulations \cite{Aoki:2008tq}, which could also be
done at $\mu\ne0$ using the overlap operator \cite{Bloch:2006cd}) is
whether the sign problem becomes harder or milder as a function of the
topological charge $\nu$.  In \cite{Bloch:2008cf} we have generalized
the results of \cite{Splittorff:2007ck} to nonzero $\nu$.  In this
contribution, we summarize our main results and present some
additional material.

Before doing so, let us briefly mention important work related to the
present subject.  The leading-order $p$-regime corrections to the RMT
result for the average phase were computed in
\cite{Splittorff:2007ck}.  The average phase at nonzero temperature
(but still at $\nu=0$) was computed from RMT in \cite{Han:2008xj}.
The distribution of the phase was studied in \cite{Lombardo:2009aw}
(see also \cite{Lombardo:2009jq} at this conference), and lattice studies of
the phase were performed in \cite{Ejiri:2008xt,Danzer:2009dk}.

\section{Random matrix model and sketch of the calculation}

The starting point of our calculation is the non-hermitian random
matrix model \cite{Osborn:2004rf}
\begin{align}
  \label{eq:rmm}
  D(m;\mu) = 
  \begin{pmatrix}
    m & i\Phi_1 + \mu \Phi_2 \\
    i \Phi_1^\dagger + \mu \Phi_2^\dagger & m
  \end{pmatrix}
\end{align}
for the Dirac operator, where $\Phi_1$ and $\Phi_2$ are independent
complex random matrices of dimension $(N+\nu)\times N$, distributed
according to the same Gaussian distribution, $P(X)\propto\exp(-N\tr
XX^\dagger)$.  The matrix in \eqref{eq:rmm} has $|\nu|$ exact zero
modes, which allows us to identify $\nu$ with the topological charge.
We keep $\nu$ fixed and take $N\to\infty$.  Without loss of generality
we assume $\nu\ge0$; for $\nu<0$ we can simply replace $\nu$ by
$|\nu|$ in our results.

Going over to an eigenvalue representation of the random matrices
$\Phi_1$ and $\Phi_2$, the partition function of the random matrix
model with $N_f$ flavors becomes \cite{Osborn:2004rf}
\begin{align}
  Z_\nu^{N_f}(\alpha) = \int_\mathbb{C} \prod_{k=1}^{N} d^2 z_k \: 
  w^\nu(z_k, z_k^*;\alpha) \: |\Delta_N(\{z^2\})|^2  \prod_{f=1}^{N_f}
  (m_f^2-z_k^2)\:,
  \label{eq:ZNf}
\end{align}
where $\alpha = \mu^2$, $\Delta_N$ is the Vandermonde determinant,
and we have introduced the weight function
\begin{align}
  w^\nu(z,z^*;\alpha) = |z|^{2\nu+2}
  \exp\left(-\frac{N(1-\alpha)}{4\alpha}(z^2+{z^*}^2)\right)
  K_\nu\left(\frac{N(1+\alpha)}{2\alpha} |z|^2\right)
  \label{eq:weight}
\end{align}
that includes a modified Bessel function $K$.  The ensemble average of
an observable is given in the usual way by including it in the
integrand and dividing the resulting integral by $Z$.

The phase factor of the squared determinant can be written as
\begin{align}
  e^{2i\theta} = \frac{\det(D(\mu)+m)}{\det(D^\dagger(\mu)+m)}
  = \prod_{k=1}^{N} \frac{m^2-z_k^2}{m^2-{z_k^*}^2}
\end{align}
with ensemble average
\begin{align}
  \ph_{N_f}
  &=\left\langle
    \frac{\det(D(\mu)+m)}{\det(D^\dagger(\mu)+m)}\right\rangle_{\!\!N_f}
  = \frac{Z^{N_f+1|1^*}_\nu\!\!(\alpha,m)}{Z^{N_f}_\nu(\alpha)}
  \notag\\
  &= \frac{1}{Z^\Nf_\nu(\alpha)}\int_\mathbb{C}
  \prod_{k=1}^{N} d^2 z_k \,  
  w^\nu(z_k, z_k^*;\alpha) \, |\Delta_N(\{z^2\})|^2
  \frac{m^2-z_k^2}{m^2-{z_k^*}^2} {\prod_{f=1}^\Nf (m_f^2-z_k^2)}\:,
  \label{eq:av}
\end{align}
where $Z_\nu^{N_f+1|1^*}$ is the RMT partition function with $N_f+1$
fermionic quarks and one conjugate bosonic quark.  The integrals in
\eqref{eq:ZNf} and \eqref{eq:av} can be done for any $N$ using the
orthogonal-polynomial formalism developed in
\cite{Akemann:2004zu,Bergere:2004cp}.  The resulting calculation is
too extensive to be reproduced here, and we therefore refer to
\cite{Bloch:2008cf} for the details.  The main ingredients are
orthogonal polynomials \cite{Osborn:2004rf} with respect to the weight
function \eqref{eq:weight} and their Cauchy transforms.  After taking
the microscopic limit with the RMT scaling variables $\hm=2Nm$ and
$\ha=2N\alpha$ (corresponding to the physical scaling variables in
\eqref{eq:scaling}) kept fixed, one arrives at
\begin{align}
  \phs_{N_f} 
  &=  \frac{1}{(2\hm)^\Nf \Nf!}
  \frac{ \begin{vmatrix}
      \I{0} & \I{1} & \cdots & \I{\Nf+1}\\[1mm]
      I_{\nu,0}(\hm) & I_{\nu,1}(\hm) & \cdots &
      I_{\nu,\Nf+1}(\hm)\\[1mm]
      I'_{\nu,0}(\hm) & I'_{\nu,1}(\hm) & \cdots &
      I'_{\nu,\Nf+1}(\hm)\\[-1mm]
      \vdots & \vdots & \vdots & \vdots \\
      I^{(\Nf)}_{\nu,0}(\hm) & I^{(\Nf)}_{\nu,1}(\hm) & \cdots &
      I^{(\Nf)}_{\nu,\Nf+1}(\hm)
    \end{vmatrix} }
  {\begin{vmatrix}
      I_{\nu,0}(\hm) & I_{\nu,1}(\hm) & \cdots &
      I_{\nu,\Nf-1}(\hm)\\[1mm]
      I'_{\nu,0}(\hm) & I'_{\nu,1}(\hm) & \cdots &
      I'_{\nu,N_f-1}(\hm)\\[-1mm]
      \vdots & \vdots & \vdots & \vdots \\
      I^{(\Nf-1)}_{\nu,0}(\hm) & I^{(\Nf-1)}_{\nu,1}(\hm) & \cdots
      & I^{(\Nf-1)}_{\nu,N_f-1}(\hm)
    \end{vmatrix} } \,,
\end{align}
where the subscript $s$ labels the microscopic limit, we have assumed
degenerate quark masses, $I_{\nu,k}(z) = z^k I_{\nu+k}(z)$ with
modified Bessel function $I$, and
\begin{align}
  \I{k} = -\frac{e^{-2\ha}}{4\pi\ha\hm^\nu}
  \int_\mathbb{C} &\frac{d^2z}{z^2-\hm^2}
  \frac{|z|^{2(\nu+1)}}{{z^*}^{\nu}} 
  \exp\left(-\frac{z^2+{z^*}^2}{8\hat\alpha}\right)
  K_\nu\left(\frac{|z|^2}{4\hat\alpha}\right) I_{\nu,k}(z^*) 
\end{align}
corresponds to the microscopic limit of the Cauchy transform mentioned
above.  The calculation of this integral is a nontrivial exercise in
complex analysis and results in
\begin{align}
  \I{k} = \frac{e^{-2\ha-\frac{\hm^2}{8\ha}}}{4\ha} & \Bigg[
  \int_0^\infty du\,
  (-)^{k} u^{k+1}
  \exp\left(-\frac{u^2}{8\ha}\right)
  I_\nu\left(\frac{\hm u}{4\hat\alpha}\right) K_{\nu+k}(u) \notag\\
  &\quad + \int_0^{\hm} du \,
  u^{k+1} \exp\left(-\frac{u^2}{8\ha}\right)
  K_\nu\left(\frac{\hm u}{4\hat\alpha}\right) I_{\nu+k}(u) \Bigg]
  + \Delta_{\nu,k}(\ha,\hm) 
  \label{eq:Hs}
\end{align}
with 
\begin{align}
  \Delta_{\nu,k}(\ha,\hm) 
  &= e^{-2\ha-\frac{\hm^2}{8\ha}} \frac{ (-)^{k}
    2^{\nu+k-1}}{\hm^\nu}  
  \sum_{i,j=0}^{i+j\leq\nu-1}
  \frac{(\nu-1-i)!(\nu+k-1-j)!}{(\nu-1-i-j)!i!j!}  
  \left( 2\ha\right)^j \left(\frac{\hm^2}{8\ha}\right)^{i}\:.
\end{align}
The $\nu=0$ limit of \eqref{eq:Hs} agrees with the result obtained in
\cite{Splittorff:2007ck}, and $\Delta_{\nu,k}$ is a new term that was
not present for $\nu=0$.  This essentially completes the calculation.

From the above equations one can derive a number of interesting
limits, such as the chiral or the thermodynamic limit.  In the latter
limit, we take $\hm$ and $\ha$ to infinity keeping their ratio fixed
and obtain
\begin{align}
  \phs^\text{th} =
  \begin{cases}
    \left(1-2\ha/\hm\right)^{\Nf+1} & \text{for } 2\ha<\hm\,,\\
    0 & \text{for } 2\ha>\hm\,,
  \end{cases}
  \label{eq:thermo}
\end{align}
where the condition $2\ha<\hm$ corresponds to $\mu<m_\pi/2$.  Note
that \eqref{eq:thermo} is independent of $\nu$ and agrees with the
result of \cite{Splittorff:2006fu,Splittorff:2007ck}.

\section{Results and discussion}

\begin{figure}
  \centering
  \includegraphics[width=0.31\textwidth]{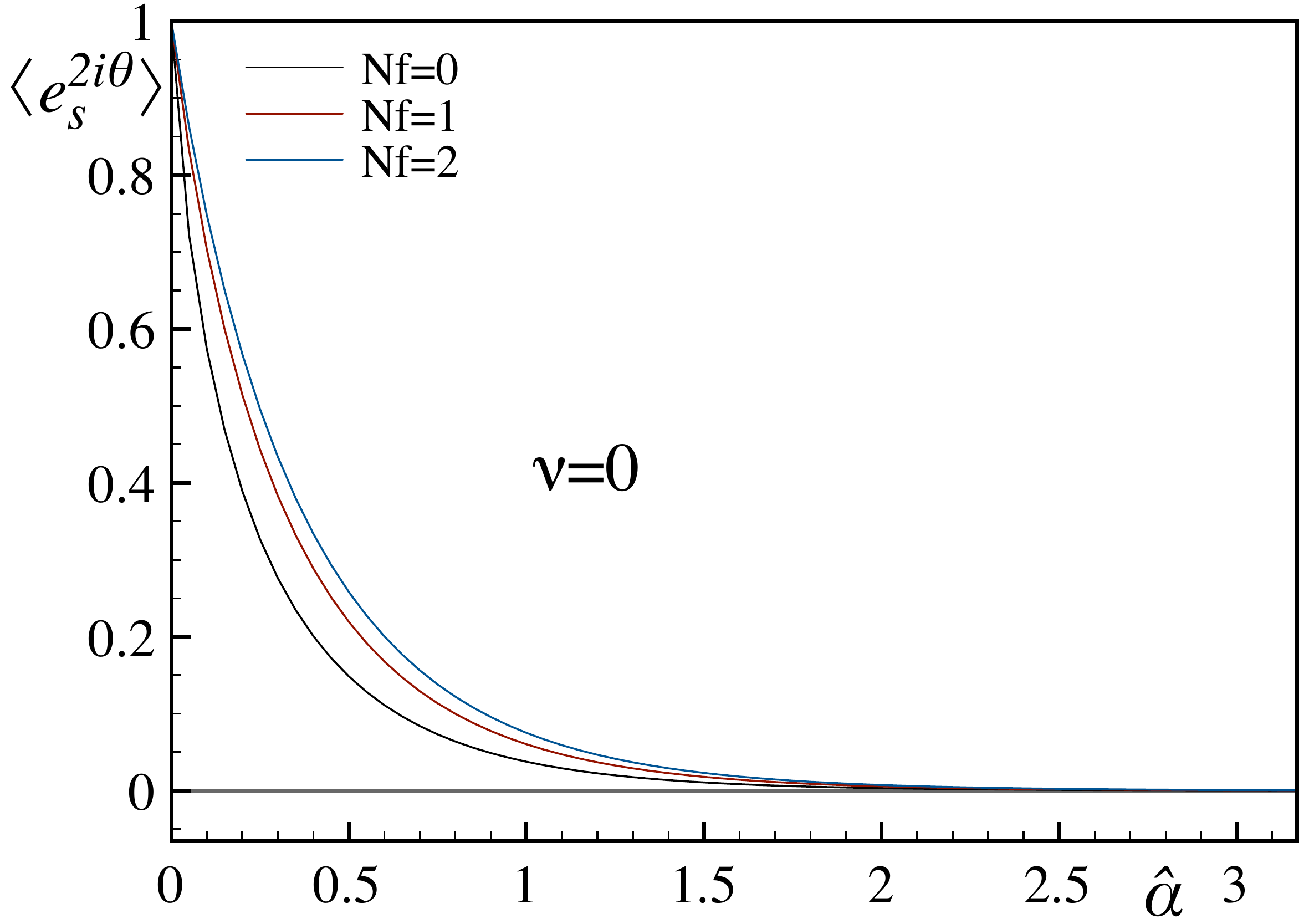}\hfill
  \includegraphics[width=0.31\textwidth]{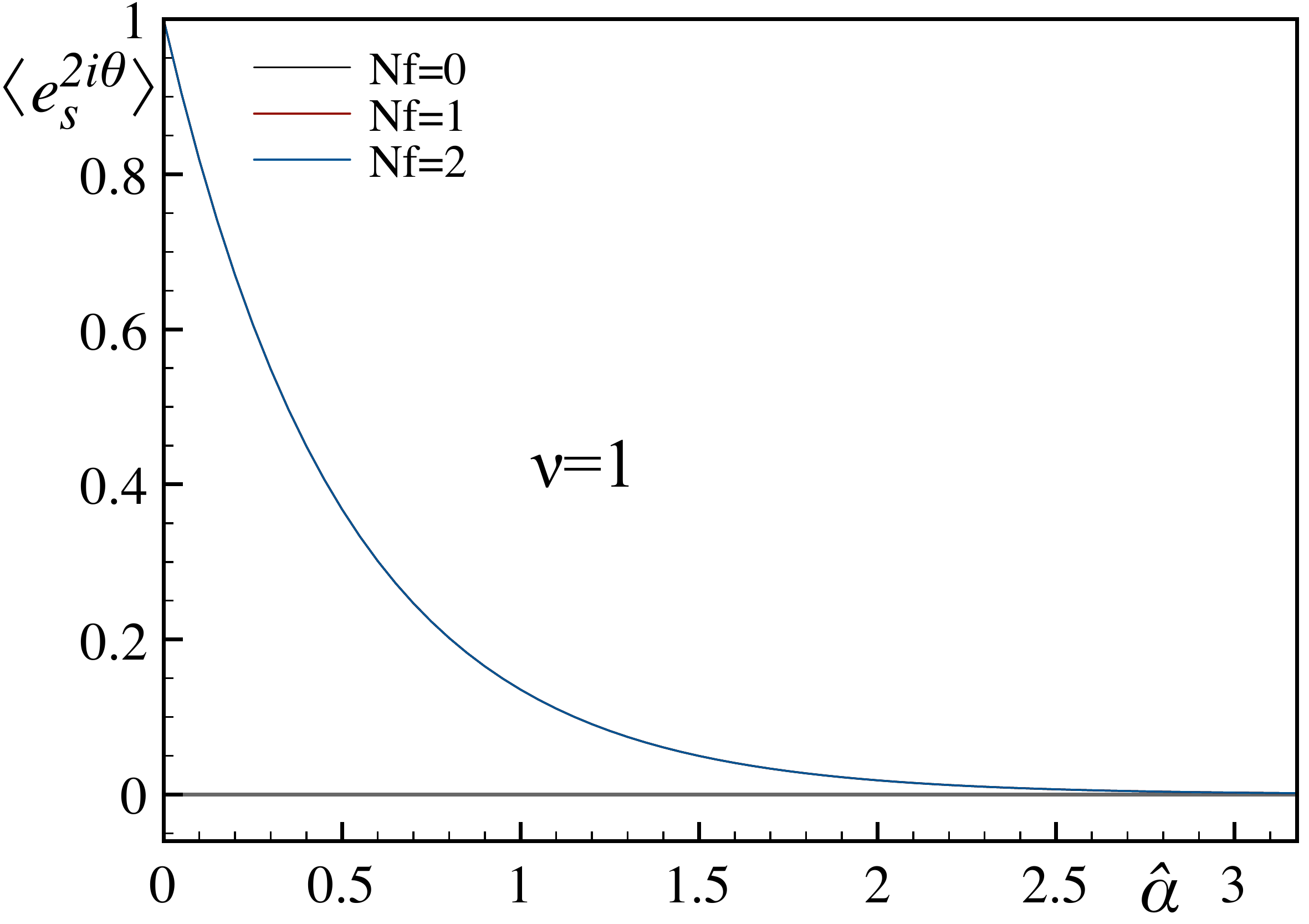}\hfill
  \includegraphics[width=0.31\textwidth]{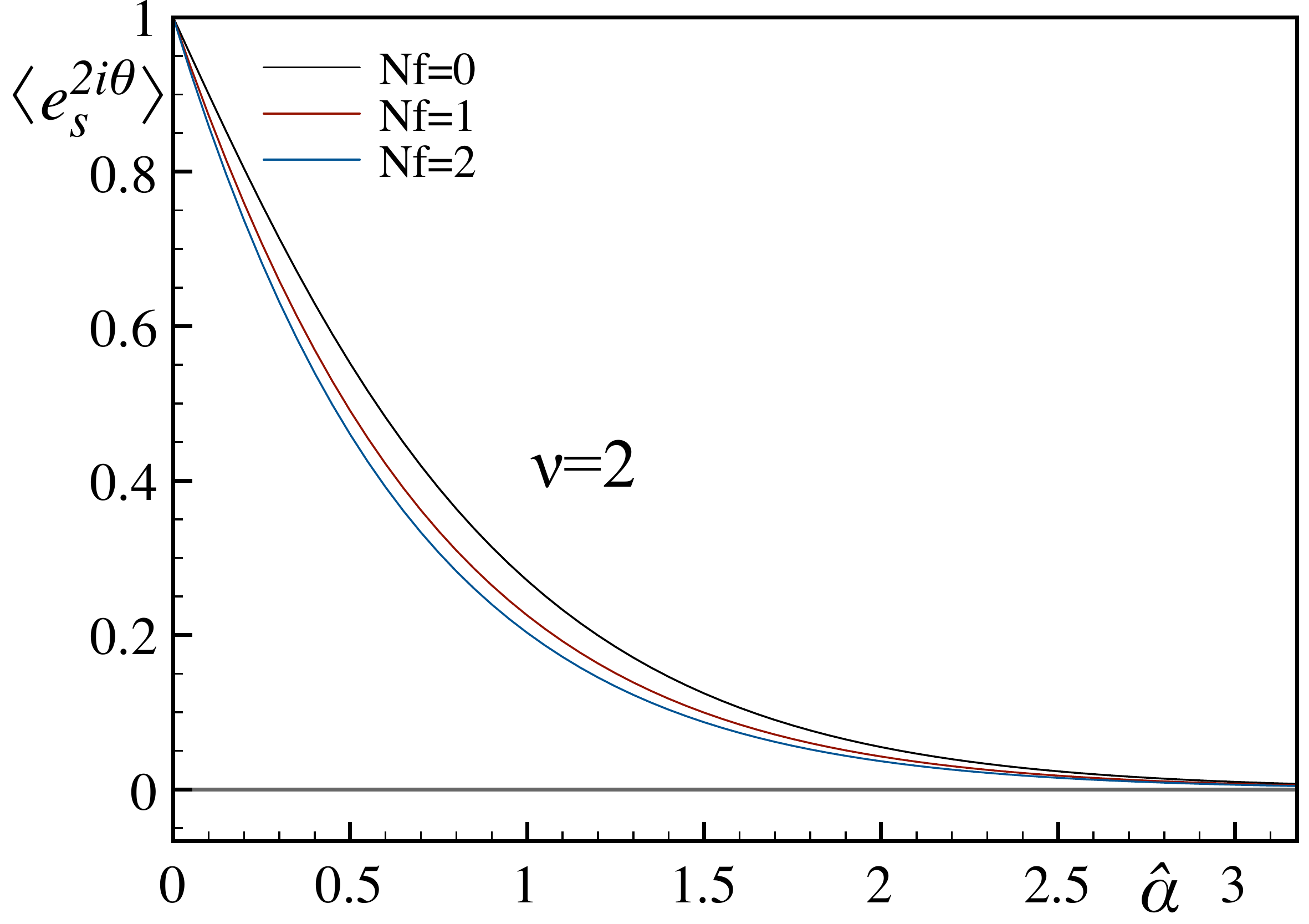}\\[3mm]
  \includegraphics[width=0.31\textwidth]{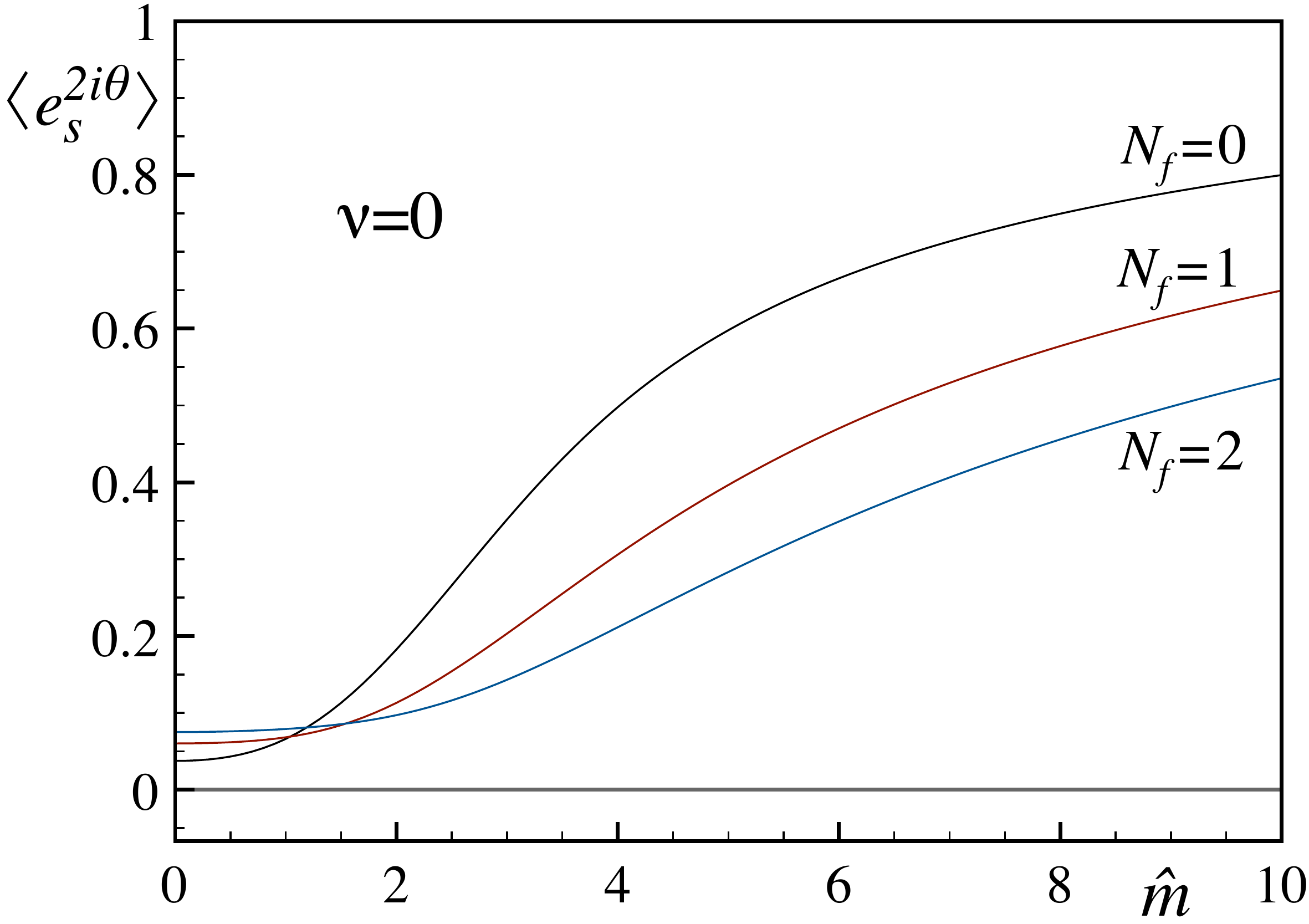}\hfill
  \includegraphics[width=0.31\textwidth]{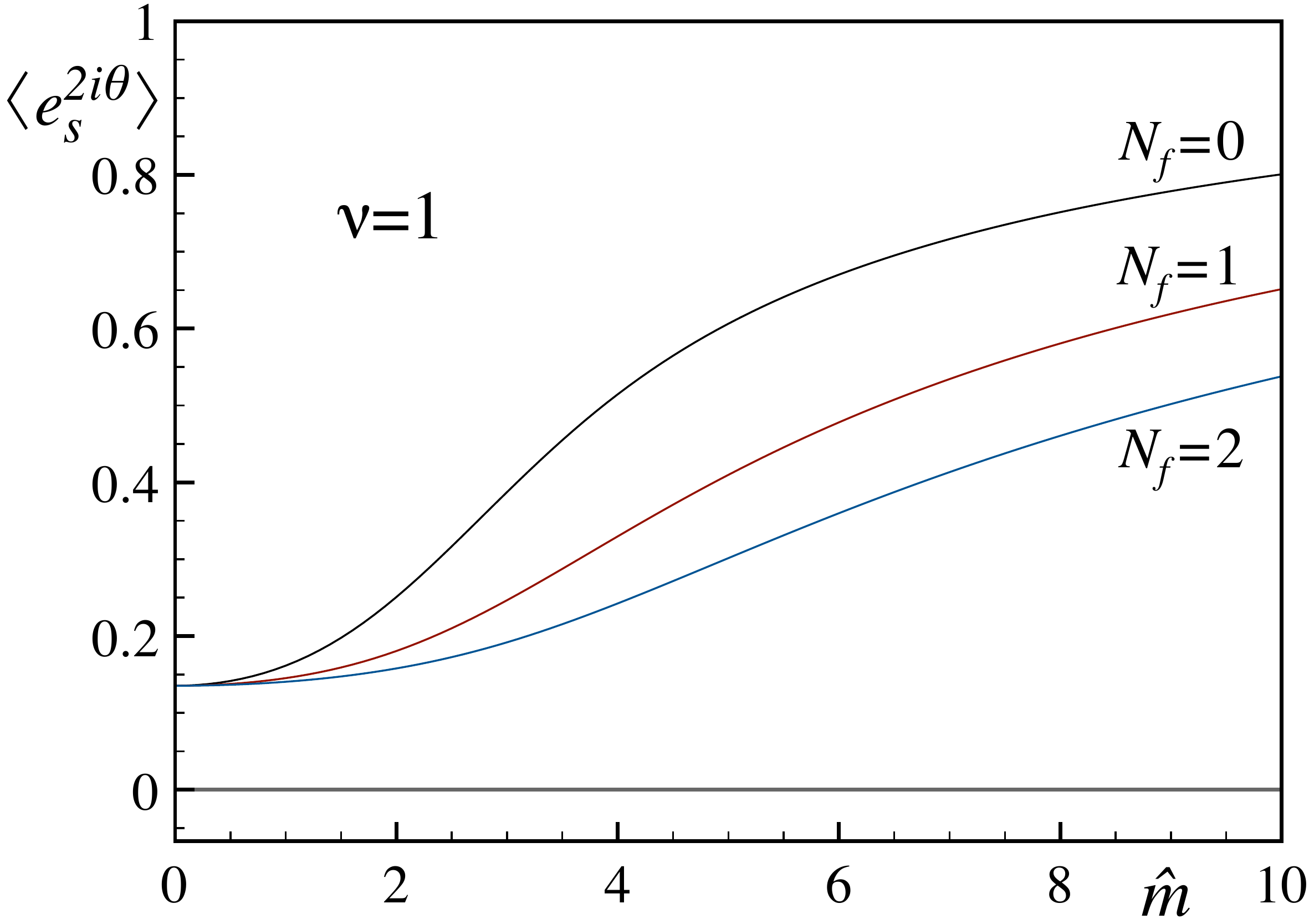}\hfill
  \includegraphics[width=0.31\textwidth]{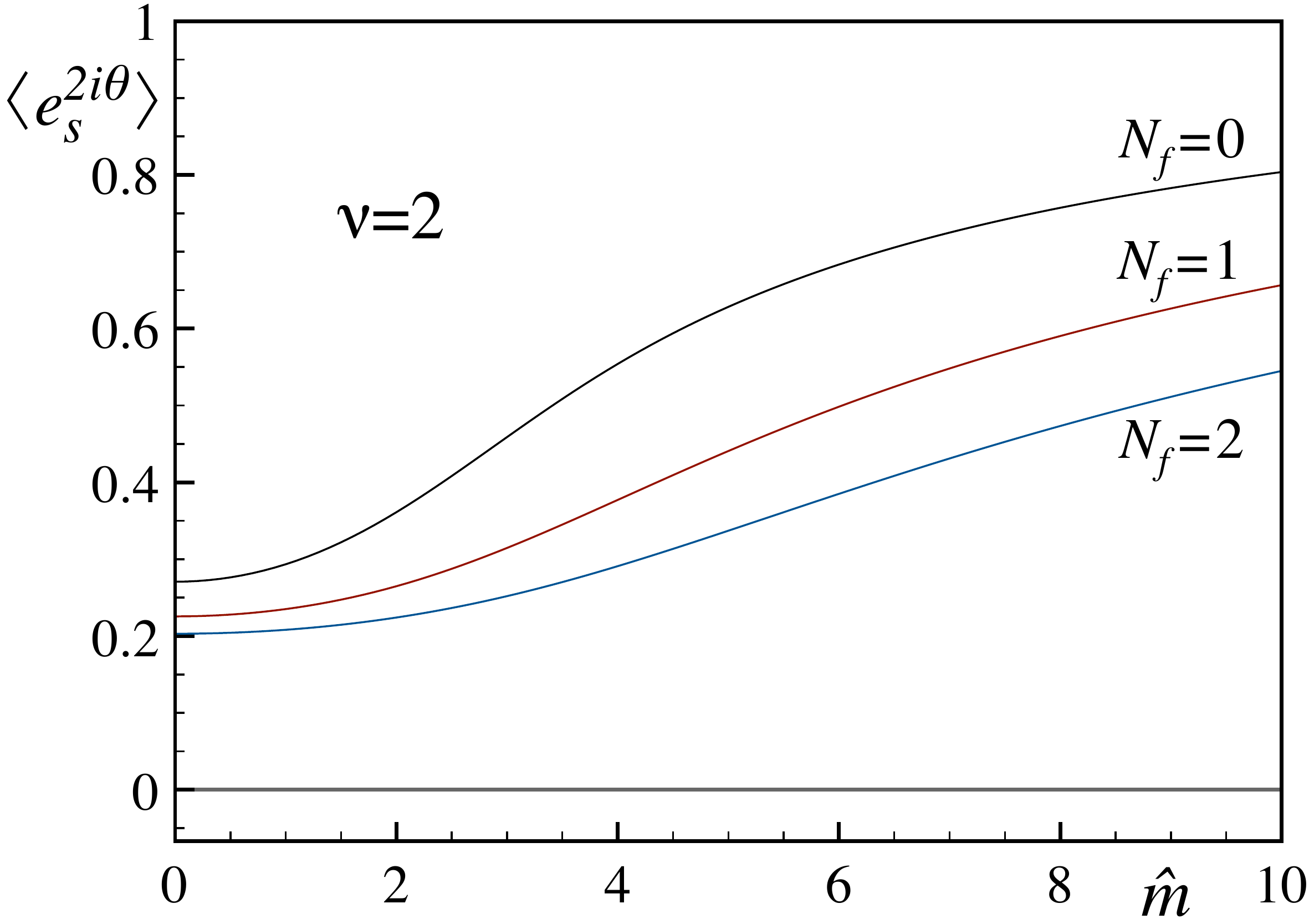}
  \caption{The top row shows the $\ha$-dependence of the average phase
    in the chiral limit for $\nu=0,1,2$ and $N_f=0,1,2$.  The bottom
    row shows the dependence on $\hm$ for fixed $\ha=1$ and the same
    values of $\nu$ and $N_f$.}
  \label{fig:results}
\end{figure}

In Fig.~\ref{fig:results} we display the dependence of the average
phase in the microscopic limit on the parameters of the problem.  As
expected, the sign problem becomes harder as we increase $\ha$ (i.e.,
$\mu$) and decrease $\hm$ (i.e., $m$). 

The dependence of $\phs$ on $\nu$ can be anticipated based on a
qualitative argument.  At $\mu=0$ the Dirac eigenvalues are purely
imaginary, and at $\mu\ne0$ they move away from the imaginary axis
into the complex plane.  The behavior of the small nonzero Dirac
eigenvalues is known from RMT \cite{Osborn:2004rf,Akemann:2007yj}.  On
average, the imaginary part of the small eigenvalues increases with
$|\nu|$, whereas the real part is mainly controlled by $\mu$ and
depends only weakly on $\nu$.  Therefore, we expect the average phase
to increase with $|\nu|$.  This expectation is confirmed by the
analytical results displayed in Fig.~\ref{fig:results}, i.e., the sign
problem becomes milder as $|\nu|$ is increased (as expected, this
effect weakens if $\hm$ is increased).  This may be an interesting
observation for groups doing lattice simulations at fixed topology in
the $\eps$-regime of QCD \cite{Fukaya:2007fb}.  As long as the
parameter $\hmu$ in \eqref{eq:scaling} is not too large, unquenched
simulations using reweighting techniques (see Sec.~\ref{sec:numerics}
below) are feasible.  It would then be advantageous to simulate not at
$\nu=0$ but at larger values of $\nu$.

As for the dependence on $N_f$, one would naively expect that more
flavors make the sign problem harder.  However, there is a competing
effect, again based on the average behavior of the small eigenvalues
known from RMT: For small $\hm$, $N_f$ has the same effect on them as
$|\nu|$, i.e., it increases the imaginary part but does not
significantly affect the real part.  As above, this competing effect is washed
out if $\hm$ is increased.  The $N_f$-dependence of $\phs$ in
Fig.~\ref{fig:results} is consistent with this qualitative argument.
We observe that for $\nu=0$ and small $\hm$, increasing $N_f$ actually
makes the sign problem milder.  Interestingly, for $|\nu|=1$ and in
the chiral limit, $\phs$ is completely independent of $N_f$.  For
$\nu>0$ the competing effect of $N_f$ does not dominate, i.e.,
increasing $N_f$ makes the sign problem harder as naively expected.

\section{Numerical random matrix simulations}
\label{sec:numerics}

\begin{figure}
  \centering
  \vspace*{5mm}
  \includegraphics[width=0.4\textwidth]{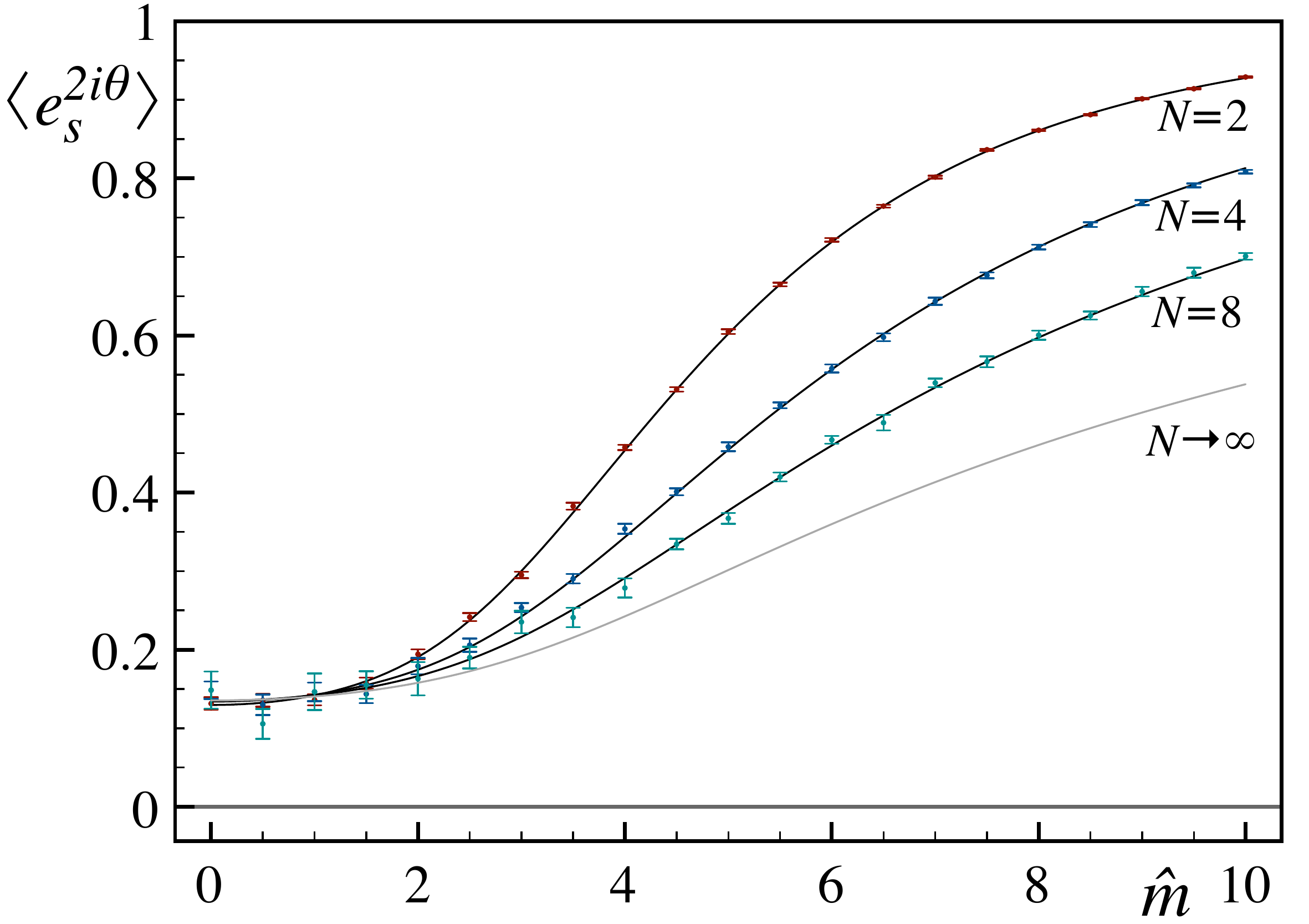}\hspace*{15mm}
  \includegraphics[width=0.4\textwidth]{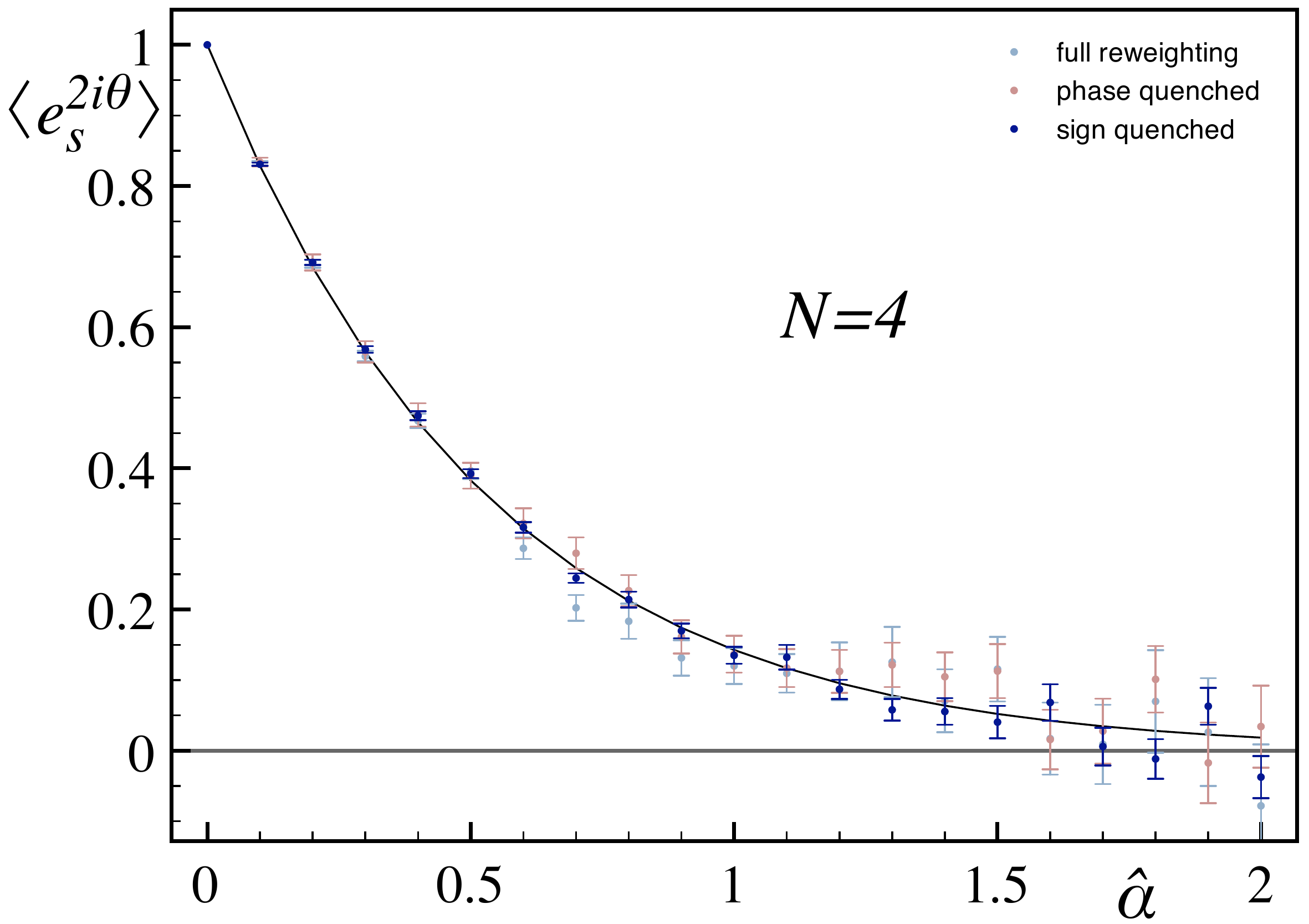}
  \caption{Dependence of the average phase on $\hm$ at fixed $\ha=1$
    (left) and on $\ha$ at fixed $\hm=1$ (right), both for $N_f=2$ and
    $\nu=1$.  The solid lines are the analytical results at finite $N$
    \cite{Bloch:2009tbp}, and the data points are from reweighted RMT
    simulations at the same value of $N$.  In the left plot
    sign-quenched reweighting was used.}
  \label{fig:unquenched}
\end{figure}

We have confirmed our analytical results (including finite-$N$
effects) by extensive numerical simulations of random matrices
\cite{Bloch:2008cf}.  Examples are shown in Fig.~\ref{fig:quenched}
(for the quenched case) and in Fig.~\ref{fig:unquenched} (for
$N_f=2$).  In the unquenched case, the numerical data were obtained by
three different reweighting methods:
\begin{figure}[h]
  \centering
  \includegraphics[width=0.31\textwidth]{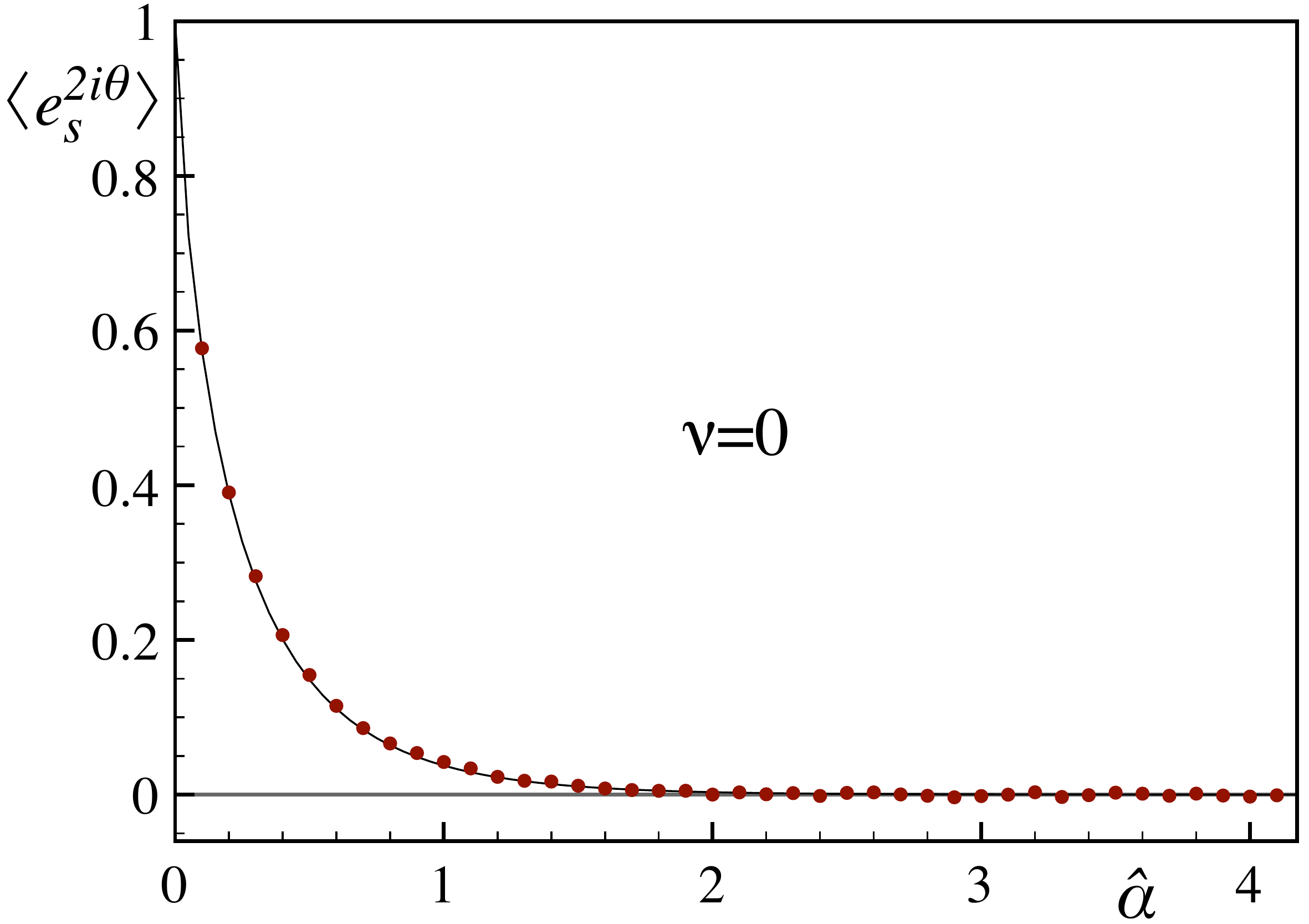}\hfill
  \includegraphics[width=0.31\textwidth]{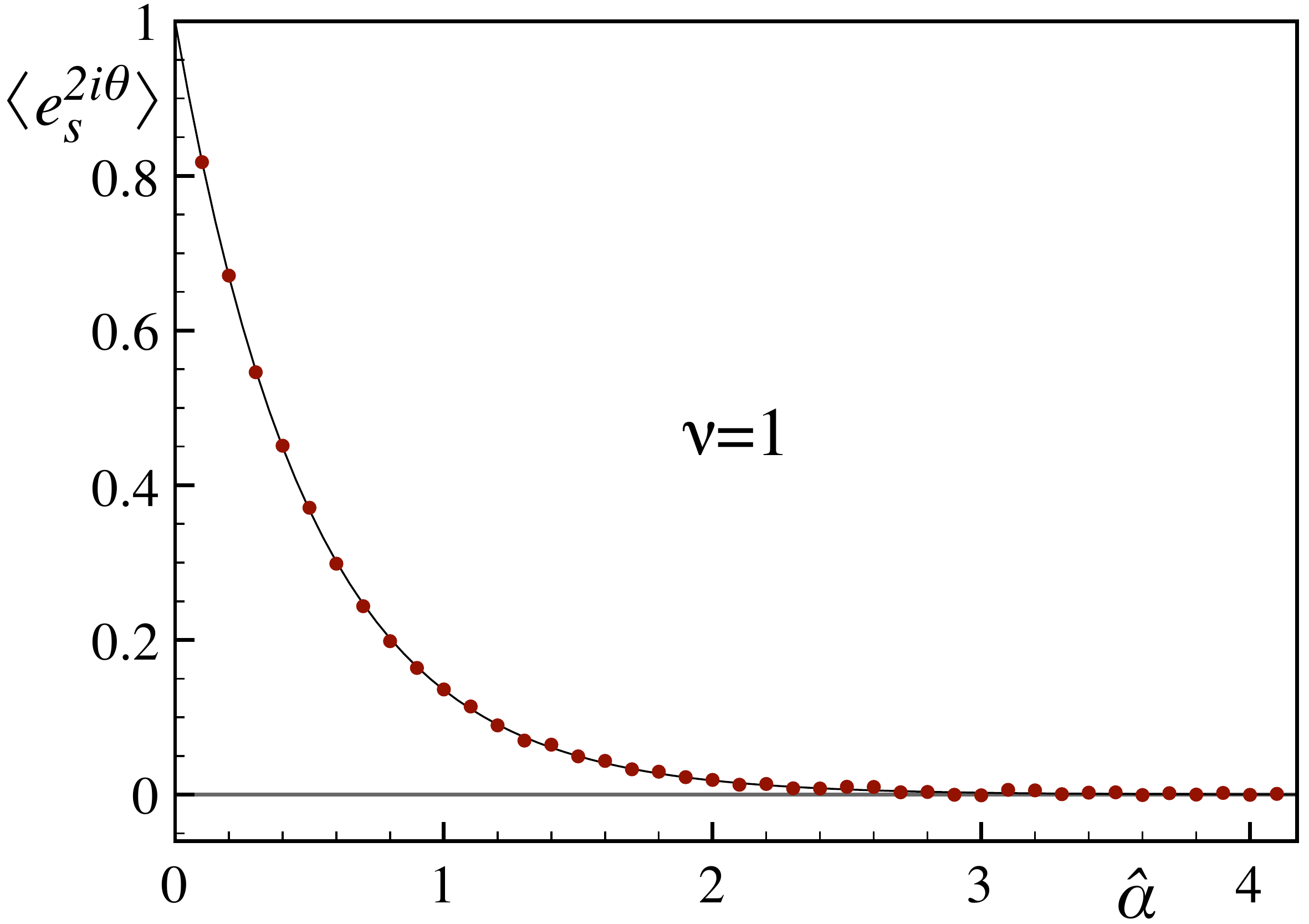}\hfill
  \includegraphics[width=0.31\textwidth]{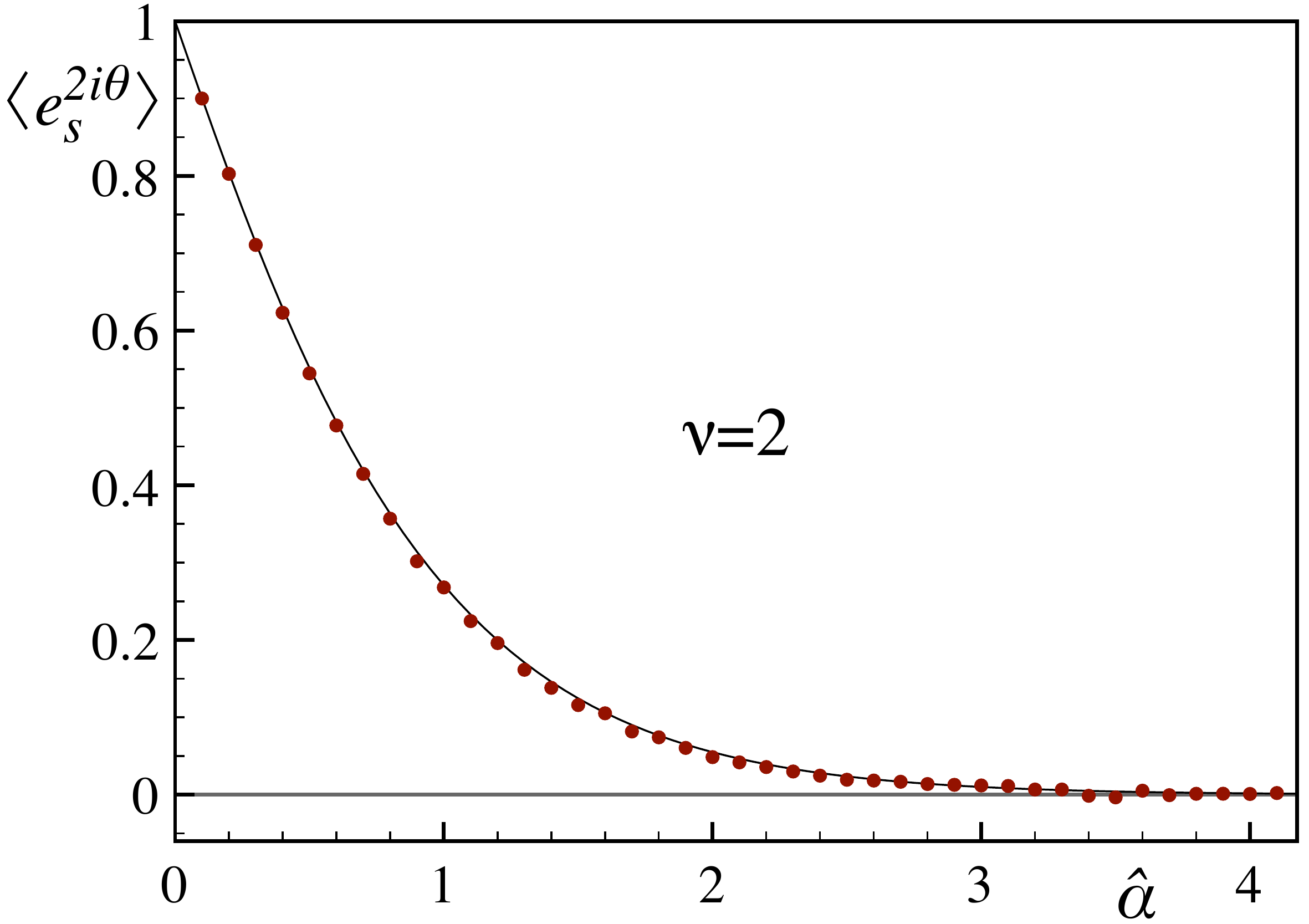}\\[3mm]
  \includegraphics[width=0.31\textwidth]{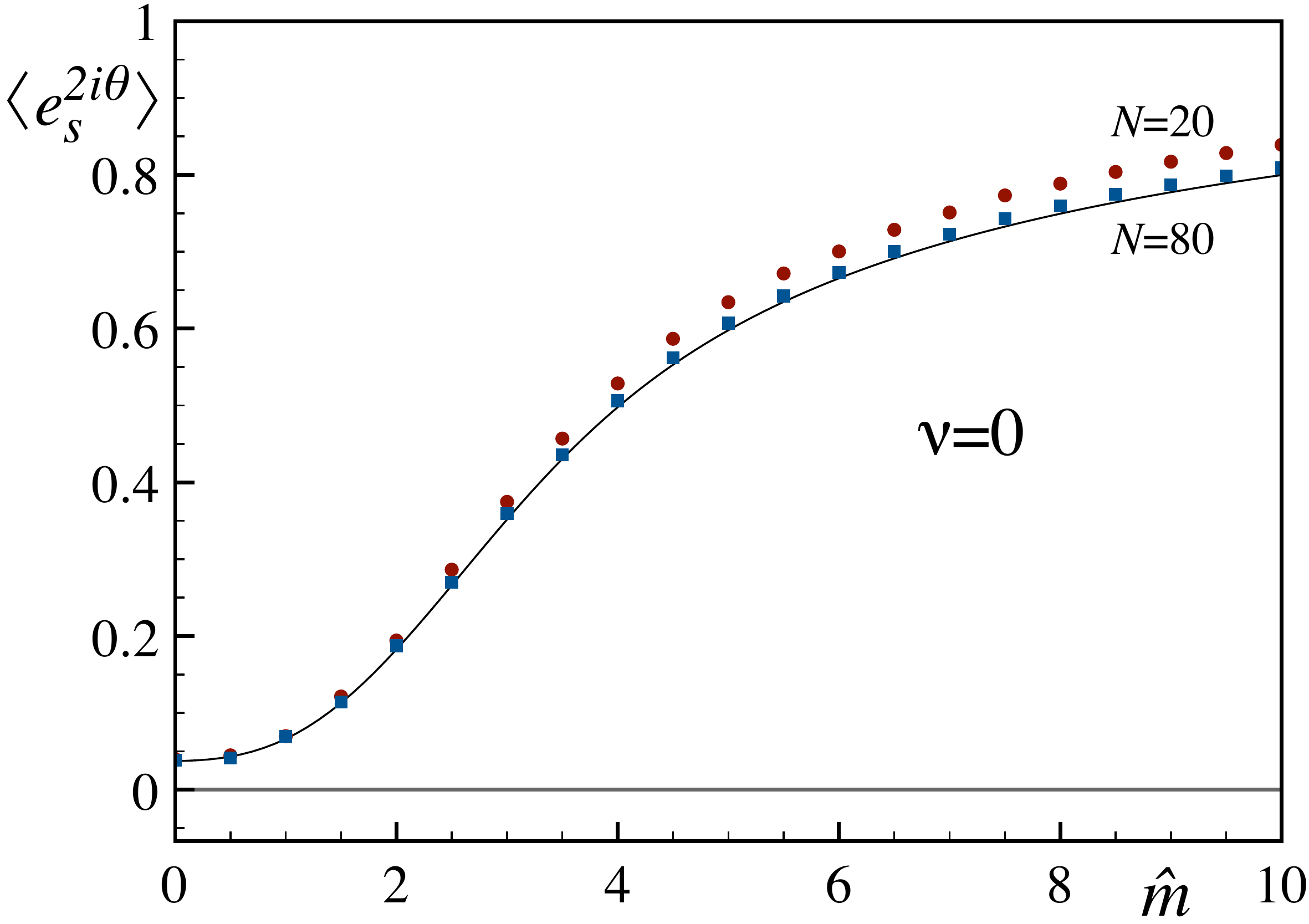}\hfill
  \includegraphics[width=0.31\textwidth]{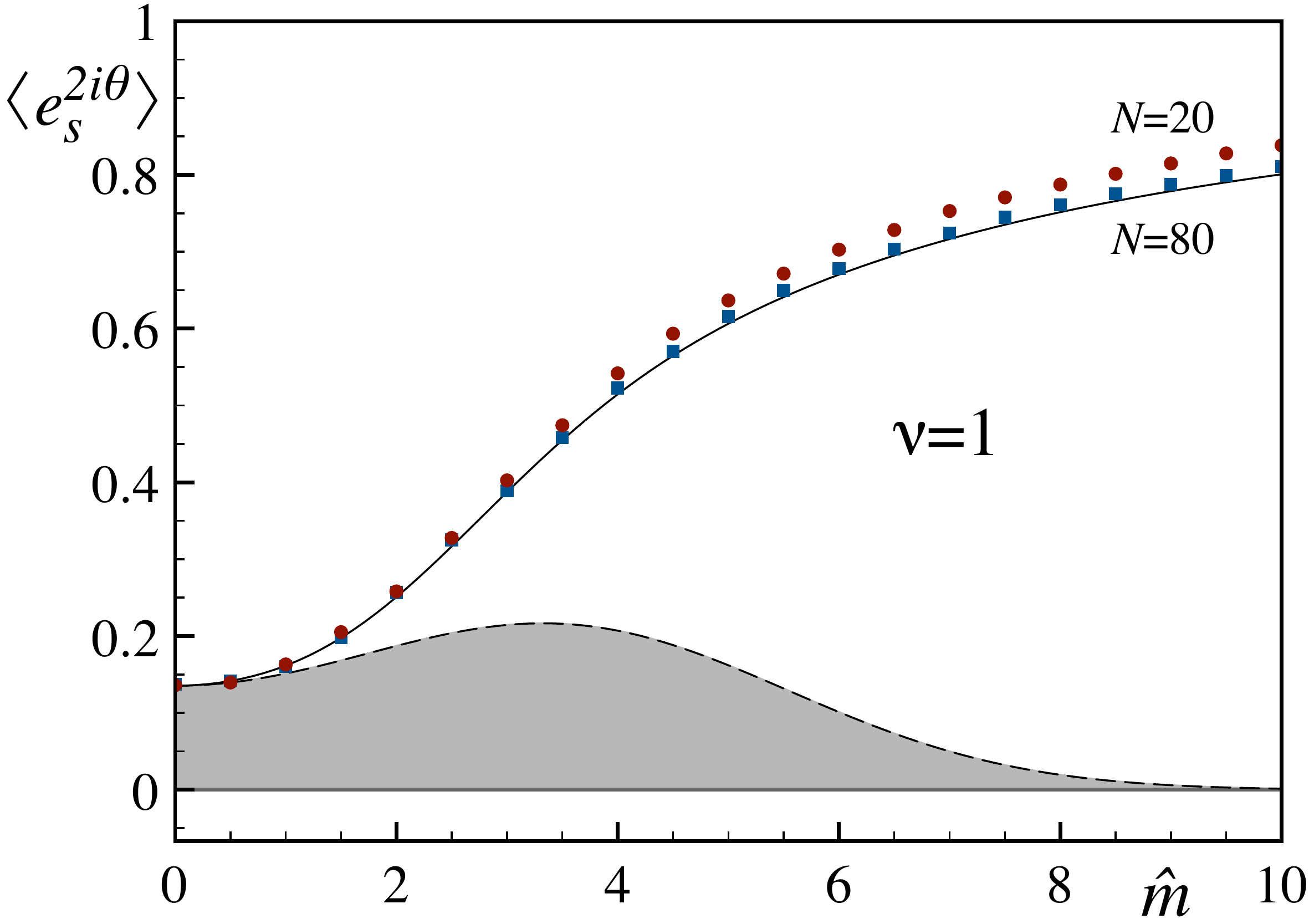}\hfill
  \includegraphics[width=0.31\textwidth]{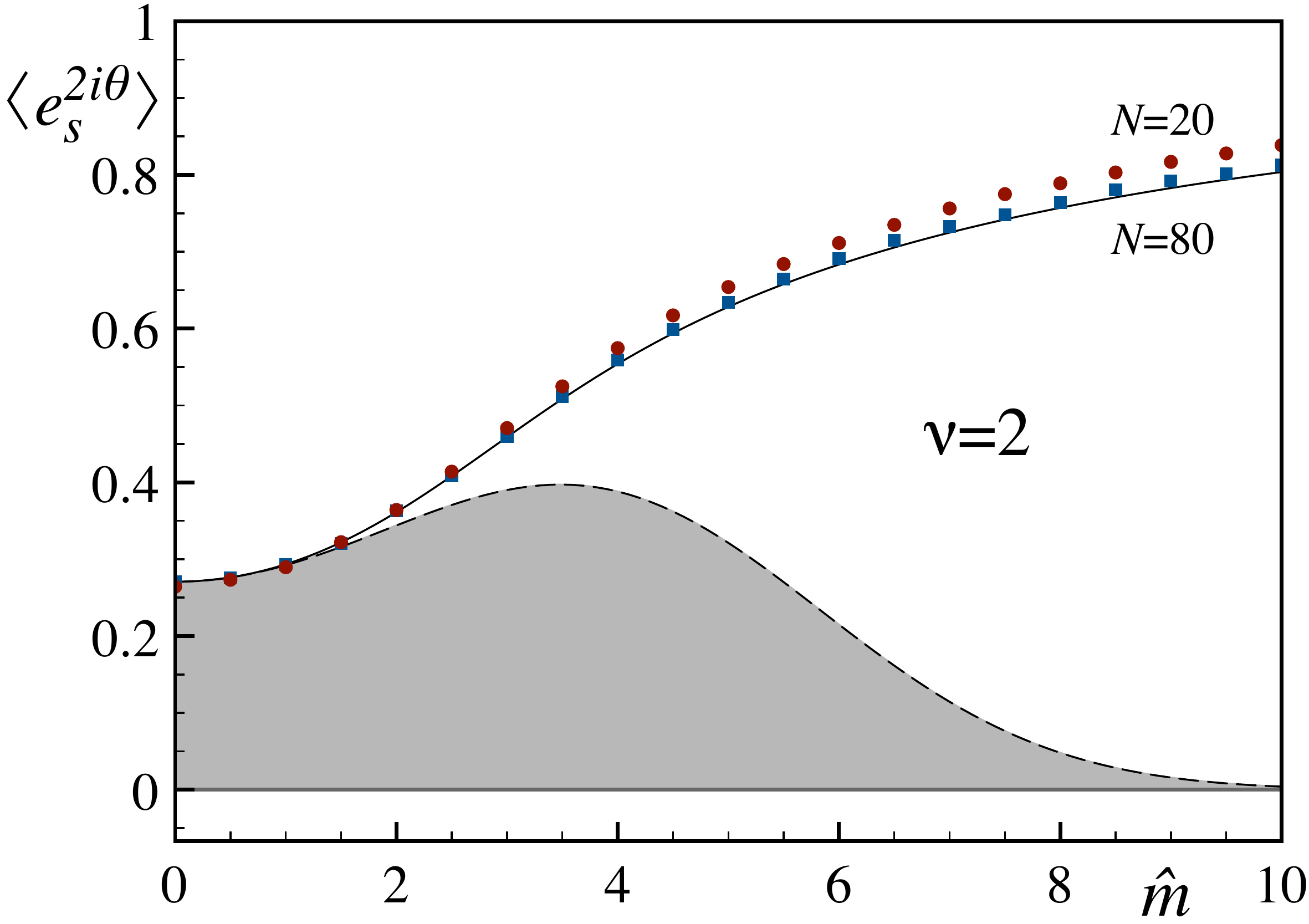}
  \caption{This figure is for the quenched case ($N_f=0$).  The top
    row shows the $\ha$-dependence of the average phase in the chiral
    limit for $\nu=0,1,2$.  The solid lines are the RMT results for
    $N\to\infty$, and the data points are from RMT simulations at
    $N=20$.  The bottom row shows the dependence on $\hm$ for fixed
    $\ha=1$, with RMT simulations at $N=20$ and $N=80$.  The
    deviations between the analytical results and the data points are
    simply finite-$N$ effects.  The filled area represents the
    contribution of the $\Delta$-term specific to $\nu \neq 0$.}
  \label{fig:quenched}
\end{figure}
\begin{enumerate}\itemsep-1mm
\item reweighting from the quenched ensemble,
\item reweighting from the phase-quenched ensemble (with $r=|\det
  D(m;\mu)|$ in the measure), 
\item\label{item:optimal} reweighting from the ``sign-quenched''
  ensemble (with $r|\cos\theta|$ and $r|\sin\theta|$ in the measure).
\end{enumerate}
Method \ref{item:optimal} minimizes the amount of reweighting that is
being done and is therefore expected to lead to the smallest
statistical errors.  This expectation is confirmed in the right plot
in Fig.~\ref{fig:unquenched}.

It would now be interesting to compare our analytical results with
data from lattice simulations.  We have done initial lattice studies,
but our volume was too small to find agreement (more precisely, we did
not have enough eigenvalues in the microscopic regime).  In the future
we hope to be able to go to larger volumes to confirm the validity of
the RMT results for QCD.

\section{Summary and outlook}

We have computed analytical results for the average phase of the
fermion determinant from RMT and found that the sign problem becomes
milder with increasing $|\nu|$.  Our analytical results are valid for
QCD in the microscopic regime.  We have confirmed these results by
extensive numerical simulations of RMT.  Interesting projects for the
future are the calculation of the distribution of the phase for
arbitrary topology, lattice simulations at larger volume, and a study
of the applicability of the sign-quenched algorithm (item
\ref{item:optimal} above) in other problems.

\bibliographystyle{JHEP}
\bibliography{tilo}

\end{document}